\documentclass{article}

\usepackage{arxiv}

\usepackage[utf8]{inputenc}
\usepackage[T1]{fontenc}
\usepackage{url}
\usepackage{booktabs}
\usepackage{amsfonts}
\usepackage{amsmath,amssymb,bm}
\usepackage{nicefrac}
\usepackage{microtype}
\usepackage{graphicx}
\usepackage{subcaption}
\usepackage{braket}
\usepackage{float}
\usepackage{threeparttable}
\usepackage{tablefootnote}
\usepackage{xcolor}
\usepackage{tikz}
\usetikzlibrary{arrows.meta}
\usepackage[font=small]{caption}
\usepackage[numbers,sort&compress]{natbib}
\usepackage{hyperref}

\renewcommand{\headeright}{Preprint}
\renewcommand{\undertitle}{Preprint}

\title{Proximal-Only Transmission Matrix Recovery of an\\
Arbitrarily Deformed Graded-Index Multimode Fiber}

\renewcommand{\shorttitle}{Proximal-Only TM Recovery in Deformed Multimode Fiber}

\author{%
  Cole Reynolds\\
  Weyl Labs\\
  \texttt{Cole@WeylLabs.com}
}
\date{}

\begin{document}

\maketitle

\begin{abstract}
The multimode fiber is among the thinnest imaging conduits available, carrying hundreds to thousands of spatial modes through a cross-section comparable to a human hair, but its endoscopic capabilities are currently limited by the sensitivity of the transmission matrix to the fiber's deformed state. Proximal-only recovery of the fiber's transmission matrix is an appealing approach for enabling general use multimode fiber endoscopy,
and within the last decade, machine learning techniques have been applied to both single-ended and double-ended transmission matrix recovery tasks. We present a new approach to this interdisciplinary problem and show that neural networks can generalize to recover transmission matrices of an arbitrarily deformed graded-index multimode fiber from proximal measurements alone.
\end{abstract}

\section{Introduction}

A Multimode fiber (MMF) propagates information between its proximal and distal ends through the reciprocal relationship

\begin{equation}
    \bm{\psi}_\text{dist} = U\bm{\psi}_\text{prox},\qquad  \bm{\psi}_\text{prox} = U^T\bm{\psi}_\text{dist}
    \label{eq:TM}
\end{equation}

where \(U\) is the fiber's transmission matrix (TM). In MMF endoscopy, the fiber's distal tip is guided to a scene described by an unknown scattering matrix \(S\), and probe signals from the proximal end propagate through the fiber, interact with the scene, and a portion of that initial signal energy propagates back through the fiber to the proximal end which can then be measured: 
\begin{equation}
    \bm{\psi}_\text{measured} = U^TSU\bm{\psi}_\text{probe}.
\end{equation}

For a known \(N\times N\) TM, the available scene content \(S\) can be recovered, in principle, with \(N\) linearly independent probe signals. This is the basis of MMF endoscopy, and it is demonstrably capable of resolving subcellular structure \cite{turtaev2018, vasquezlopez2018, ohayon2018, wen2023}.

Despite these successes, MMF endoscopy applications are limited by knowledge of the underlying fiber's TM because the TM changes with the fiber's geometry: bends and twists during endoscopy change the TM, and since the distal tip at the scene cannot be accessed, the TM cannot be directly measured or calibrated using Eq. \eqref{eq:TM}. Every other requirement of MMF endoscopy has been achieved experimentally: probe delivery, scene interrogation, and image formulation all follow from Eq. \eqref{eq:TM} once \(U\) is known, thus recovery of the deformed fiber's TM is the bottleneck to general use. Proximal-only recovery methods \cite{gu2015} try to obtain the TM through the use of reflectors located at the distal end of the fiber. Reflection matrices within a fiber, unlike a varying and unknown scene scattering matrix outside a fiber, are constant across measurements: for \(K\) reflectors with reflection matrices \(\{R_k\}\), one probe signal provides the \(K\) measurements

\begin{equation}
    \bm{\psi}_\text{measured}^{(k)} = U^T\tilde R_{k}U\bm{\psi}_\text{probe} = M_k\bm{\psi}_\text{probe},
\end{equation}

Where \(\tilde R_i\) denotes the interaction with \(R_i\) as well as any prior or subsequent reflection, transmission, or propagation matrices.

This approach is appealing because it re-frames the historically physics problem as a machine learning problem: given a set of measurements \(\{M_k\}\), does there exist a mapping that accurately encodes the state of an arbitrarily deformed fiber, and can that encoded state be decoded to recover the TM? Deep learning's canonical successes never needed to ask this question because it already had empirical evidence of the answer: a finite nervous system performs vision and language, so those domains provably admit finite representations, and the architectures that succeeded there were built to exploit relevant abstract structure such as locality, invariance, and equivariance. A physics problem has no such guarantee. Whether a fiber's transmission matrix admits a finite parameterization under deformation is a property of the fiber, and prior works provide little evidence that the inductive biases of image-domain architectures are aligned with the fiber's dynamics. A learned model must recover the TM of a fiber state it has never encountered, and thus the model must capture and encode what the fiber does under deformation rather than what a sample of deformations looked like. In practice, the encoding must be equivariant in the informal sense that a physical action on the fiber, a bend or a twist, acts on the encoded state as a corresponding structural operation, and it is this property that allows the model to recover states it has not seen. In this work we present a new approach that suggests a finite set of graded-index (GRIN) training fibers is enough to learn that structure, and we demonstrate capabilities on a synthetic commercial grade fiber under realistic measuring conditions.

\begin{figure}[h!]
    \centering
    \includegraphics[width=0.95\textwidth]{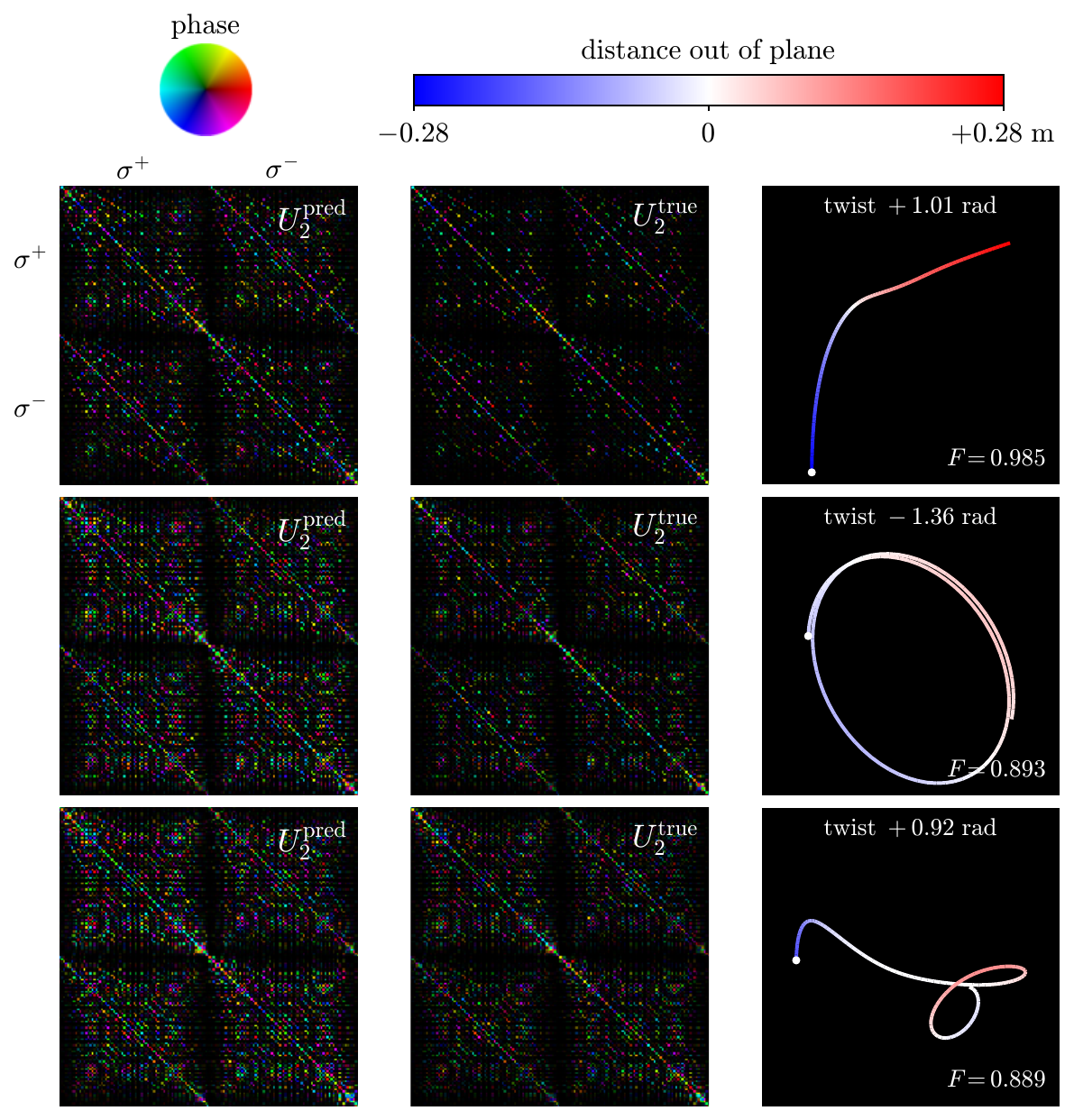}
    \caption{Architecture results for 3 held-out GRIN fibers in different recovery regimes. Only every 4th column and row are shown for visual clarity.}
    \label{fig:eval}
\end{figure}

\section{Related Work}
Currently, the most successful approach to MMF endoscopic imaging is to make the fixed-TM assumption hold: the TM is characterized before deployment, with access to both facets, and the fiber's geometry is then locked for the duration of imaging. The \v{C}i\v{z}m\'ar group has carried this paradigm the furthest. They first demonstrated lensless imaging through a standard MMF \cite{cizmar2012} and then later demonstrated high-fidelity fluorescence endoscopy deep in the living brain \cite{turtaev2018}.

The general TM deformation problem has also been addressed directly from the fiber side. Flaes et al. \cite{flaes2018} showed that light transport through parabolic-profile GRIN fibers is intrinsically more robust to bending than through step-index fibers, and predicted that sufficiently precise parabolic-index fibers would make flexible probes feasible. Subsequent work pursued bending resilience through a hybrid multimode-multicore design \cite{du2022}, and more recently through tomographic measurement of the refractive-index perturbations of commercial GRIN fibers, showing that the measured profiles account for the observed decline of imaging under bending, identifying the fibers that endure it best, and demonstrating a flexible endoscope built on one of them \cite{turtaev2025}. This approach aims to solve the TM deformation problem by making the TM sufficiently deformation-invariant. This bend-resilience narrows the gap between the calibrated TM and the true underlying TM, but the gap still grows with the severity of the deformation and cannot be accounted for without some form of active sensing. Nevertheless, tolerance aids recovery and we argue in Section \ref{sec:approach} that the mechanism behind the bend-resilience of graded-index fibers, low-dimensional parameterization, is what enables general TM recovery.

\section{Theory}
\label{sec:approach}

We first show that the TM for an arbitrarily deformed ideal GRIN fiber admits a finite representation while the TM for a step-index fiber does not.

\subsection{Graded-Index Fiber}
In the weakly-guided approximation \cite{snyderlove}, the Hamiltonian for a bent and twisted ideal GRIN fiber is the linear combination of operators (see appendix A)

\begin{equation}
    H(z) =  c_0\, p^2 + c_1 r^2 + c_2\,\bm\kappa(z)\!\cdot\!\bm r + c_3\,\tau(z)\,L_z,
    \label{eq:Hdeformed}
\end{equation}

This set of deformation operators is not closed under commutation since \([ p^2,  r^2] = -2i(\bm r\cdot \bm p + \bm p\cdot\bm r)=-2i\,D\), but the set with \(D\) included does close. The non-trivial commutation relations are

\begin{equation}
\begin{aligned}
    [x_j, p_l] &= i\,\delta_{jl}\,I, &\qquad
    [x_j,  p^2] &= 2i\,p_j, &\qquad
    [x_j, D] &= 2i\,x_j, &\qquad
    [x_j, L_z] &= -i\,\varepsilon_{jl}\,x_l, \\
    [p_j,  r^2] &= -2i\,x_j, &\qquad
    [p_j, D] &= -2i\,p_j, &\qquad
    [p_j, L_z] &= -i\,\varepsilon_{jl}\,p_l, \\
    [ p^2,  r^2] &= -2i\,D, &\qquad
    [ p^2, D] &= -4i\, p^2, \\
    [ r^2, D] &= 4i\, r^2,
\end{aligned}
\label{eq:closure}
\end{equation}

where \(\varepsilon\) is the two-dimensional Levi-Civita symbol and the remaining relations \([x_j,r^2]\), \([p_j,p^2]\), \([r^2,L_z]\), \([p^2,L_z]\), and \([D,L_z]\) all vanish. Thus, \(\mathfrak{g} = \operatorname{span}\{I, x, y, p_x, p_y, L_z,  p^2,  r^2, D\}\) is a nine-dimensional Lie algebra that contains \(H(z)\) for all deformations.

Having the general Hamiltonian contained within a finite-dimensional Lie algebra is what makes the proximal-only TM recovery problem tractable. The TM for any deformed fiber takes the path-ordered form

\begin{equation}
\label{eq:pathorder}
    U = \mathcal{P}\exp\!\left(-i\int_0^L H(z)\,dz\right).
\end{equation}

 Dividing the fiber into segments of length \(\Delta z\) over which \(H\) is constant, Eq.~\eqref{eq:pathorder} is the limit of the ordered product
\begin{equation}
    U = \lim_{\Delta z \to 0}\, e^{-iH(z_M)\Delta z} \cdots e^{-iH(z_2)\Delta z}\, e^{-iH(z_1)\Delta z},
    \label{eq:segments}
\end{equation}
and neighboring segments combine through the Baker--Campbell--Hausdorff (BCH) formula,
\begin{equation}
    e^{A}\, e^{B} \;=\; \exp\!\Big( A + B + \tfrac{1}{2}[A,B] + \tfrac{1}{12}\big([A,[A,B]] + [B,[B,A]]\big) + \cdots \Big).
    \label{eq:bch}
\end{equation}
But since \(\mathfrak{g}\) is closed, the expansion in Eq.~\eqref{eq:bch} for the GRIN fiber can always be expressed as a linear combination of elements in the Lie algebra,

\begin{equation}
    e^{A}\, e^{B} \;=\; \exp\! \left(\sum_{i}\alpha_i g_i\right),\qquad g_i\in \mathfrak{g},
    \label{eq:bch_lc}
\end{equation}

and the same holds for any pair of exponentials built from \(\mathfrak{g}\). An ordering change therefore adds no new terms and only changes the coefficients: applying Eq.~\eqref{eq:bch_lc} to both \(e^{-A}e^{-B}\) and \(e^{A}e^{B}\), and then to the product of those two results gives

\begin{equation}
\begin{aligned}
    e^{A}e^{B} &= e^{B}e^{A}\,X, \\
    X &= e^{-A}e^{-B}e^{A}e^{B}\;=\; \exp\!\left(\sum_{i}\beta_i g_i\right)\exp\!\left(\sum_{i}\gamma_i g_i\right) \;=\; \exp\!\left(\sum_{i}\delta_i g_i\right),
\end{aligned}
\label{eq:swap}
\end{equation}

where \(\bm\beta\), \(\bm\gamma\), and \(\bm\delta\) are each fixed by \(A\) and \(B\) through Eq.~\eqref{eq:bch}. Exchanging two segments therefore costs one further element of the same group rather than an expansion without end, and applying Eq.~\eqref{eq:bch_lc} across every segment of Eq.~\eqref{eq:segments} collapses the ordered product to a single exponential,
\begin{equation}
    U = f(\bm a) = \exp\!\left(\sum_{i} a_i g_i\right),\quad g_i\in\mathfrak{g},
    \label{eq:factored}
\end{equation}
where \(\bm a\) is fixed by the deformation profile \cite{magnus1954, wei1963}. The interpretation is that while the physical geometry of any fiber is path-ordered, the TM dynamics of the GRIN fiber are not, and recovery from finite measurements is possible because those dynamics have a finite parameterization.

\subsection{Step-Index Fiber}

The step-index fiber replaces the quadratic well with the canonical finite-well profile, but we assume a general potential of the form \(V(\bm r) = \sum_m c_m\, r^m\). The GRIN fiber is the special case \(c_{m\neq 2} =0\). Using the identity \([A, BC] = [A,B]\,C + B\,[A,C]\), every \(r^m\) for \(m\geq 3\) in the Hamiltonian can be generated by \(r\) and \( r^2\), but only \(m=2\) creates a finite Lie algebra: 
\begin{align}
    [p^2, r] &= \frac{1}{r}\,\left(-2i\bm r\!\cdot\!\bm p - 1\right),
    \\
    [p^2, r^4] &=  r^2\,[ p^2,  r^2] + [ p^2,  r^2]\, r^2
    \\
    &= -2i\,\left( r^2 D + D\, r^2\right).
    \label{eq:oddseed}
\end{align}
The non-quadratic Hamiltonian is

\begin{equation}
    H(z) =  c_0\, p^2 + c_1\,\bm\kappa(z)\!\cdot\!\bm r + c_2\,L_z + \sum_{m=0}^{\infty}c_{m+3}\,r^m
    \label{eq:Hstep}
\end{equation}

and performing the same analysis (Eqs.~\eqref{eq:pathorder}, \eqref{eq:segments}, and \eqref{eq:bch}) results in a TM with no finite parameterization. The BCH formula does not terminate and the path-ordering does not become tractable. The number of operators needed to describe the fiber grows with the number of bends rather than remaining fixed, and in the continuum limit, where the curvature varies smoothly along the fiber and the number of segments is unbounded, no truncation of the expansion is uniformly valid. Recovering the TM of a step-index fiber is consequently equivalent to recovering \(\bm\kappa(z)\) at every point along its length, which no finite set of proximal measurements can determine.

\subsection{Real Fibers}

For the parabolic profile, the \(z\)-dependent deformation functions \(\bm\kappa(z)\) and \(\tau(z)\) never appear in the TM individually: closure consolidates each generator's accumulated contribution into a single scalar, so the TM of Eq.~\eqref{eq:factored} depends on the deformation history only through history-dependent, weighted integrals of the deformation profile, and its dynamics are exhausted by how those nine scalars evolve. A non-parabolic profile cannot be similarly compressed because each application of Eq.~\eqref{eq:bch} in an effort to consolidate terms comes at the cost of an infinite expansion, and recovering the TM becomes tantamount to recovering the deformation functions themselves point by point along the fiber.

In practice, commercial GRIN fibers are not perfectly parabolic, and contain many defects that break this ideal structure. Fortunately, the defects in the spatial refractive profile for GRIN fibers are perturbative, and parameter consolidation introduces higher-order BCH terms that decay rapidly, and general recovery from finite measurements only needs the dynamics to be well approximated by some finite representation.




Demonstrating general recovery requires data sampled from fibers whose curvature admits no convenient parameterization and this data would be expensive to obtain experimentally. Consequently, we establish the architecture's capabilities on synthesized fibers, where deformation states are sampled from a continuum. Details about the Hamiltonian used to describe the fiber's dynamics can be found in Appendix \ref{app:hamiltonian}, and details about how the fiber is deformed and how the TM and reflection measurements are obtained can be found in Appendix \ref{app:simulator}.

\section{Methods}
\subsection{Proximal-Only Recovery System for MMF Endoscopy}
In a physically realizable MMF endoscopic system, an input signal propagates down the fiber, interacts with the reflectors and the scene, and their respective return signals are measured back at the proximal end. The \(N\times N\) reflection and scene matrices can be obtained from \(N\) linearly independent input signals using an amplitude + phase measuring technique such as off-axis holography.

\begin{figure}[h!]
\centering
\includegraphics[width=0.95\linewidth]{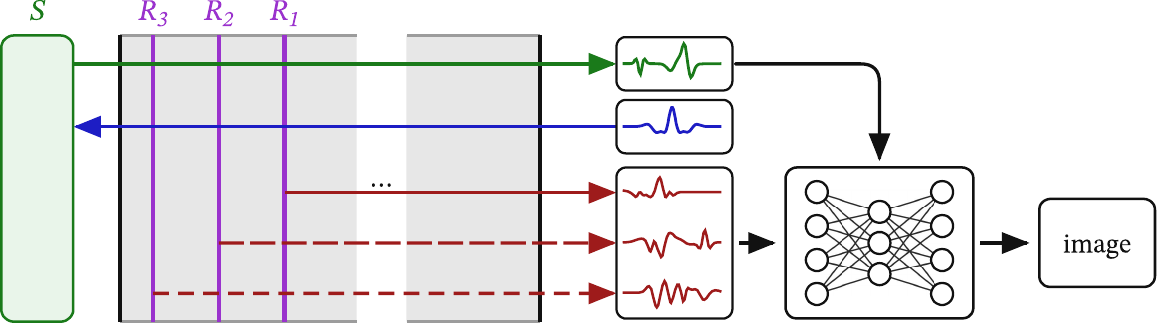}
\caption{Visual representation of our proximal-end recovery system. The input signal (blue) propagates down the fiber, reflects off of reflectors \(R_1\), \(R_2\), \(R_3\), and the scene \(S\), and then propagates back. The TM is determined from the measurement signals (red) and then uses the scene signal (green) to construct the final image.}
\label{fig:recovery-mechanism}
\end{figure}

All matrices and signals are represented in the Laguerre-Gauss spatial mode basis --- the propagation-invariant modes (PIM) for the ideal GRIN fiber. These PIMs carry a complex \(e^{i\ell\varphi}\) factor, where propagation back to the proximal end after reflection is the transpose composed with the \(\ell\to-\ell\) operator \(P\) for circularly polarized light. Writing \(\mathcal{P}[X] \equiv PXP\), the round trip reflection measurements are \(M_i = \mathcal{P}[U^{T}]\,R_i\,U\).

Since these measurements are bilinear, at least three reflectors are needed to break all symmetries and uniquely determine \(U\) \cite{gu2015, lee2020}. In this work we use a straight partial reflector \(R_1\), a tilted partial reflector \(R_2\), and a second tilted partial reflector \(R_3\) with its tilt axis making a \(\pi/4\) angle with \(R_2\)'s tilt axis. We note that neither \(R_i\) nor \(\tilde R_i\) need to be explicitly known or fabricated with extreme precision in practice. The architecture only requires that all bilinear-induced symmetries are sufficiently broken by the set of reflectors.

The architecture solves for a fiber's bent TM, \(U_2\), given the fiber's unbent TM, \(U_1\), and the three reflector measurements \(\{M_1, M_2, M_3\}\). The unknown \(U_2\) is obtained indirectly by solving for \(V = U_2U_1^\dagger\) instead, where \(U_2 = VU_1\) and \(V\) is the operator that describes the mode couplings due to the deformation. Using \(\mathcal{P}[XY]=\mathcal{P}[X]\mathcal{P}[Y]\),
\begin{align}
    M_i &= \mathcal{P}\big[(VU_1)^{T}\big]R_i(VU_1)
         = \mathcal{P}[U_1^{T}]\,\mathcal{P}[V^{T}]R_iV\,U_1
         = \mathcal{P}[U_1^{T}]\,A_i\,U_1 \\
    A_i &= \mathcal{P}[V^{T}]\,R_i\,V
         = \mathcal{P}[U_1^{*}]\,M_i\,U_1^\dagger ,
\end{align}
where \(A_i\) is a known since both \(M_i\) and \(U_1\) are known. 

\subsection{Architecture}

We use a bilinear architecture \cite{memisevic2007, reynolds2026} with one encoder per measurement \(A_i\), and replace the correspondence decoder in \cite{reynolds2026} with a MLP that outputs \(V\):
\begin{align}
    h^i_{jk} &= \bra{\bm{\phi}^i_j}A_i\ket{\bm{\psi}^i_k},
    \\
    g^i_n &= \sigma(W_{jkn}h^i_{jk} + c_n),
    \\
    \textrm{cat}[\textrm{Re}(V),\textrm{Im}(V)] &= \textrm{MLP}(\textrm{cat}[\mathbf{g}^1, \mathbf{g}^2, \mathbf{g}^3]),
\end{align}

where \(\sigma\) is a non-linear activation function (we use tanh in this paper) and the MLP contains two hidden layers with the leaky ReLU activation function (\(\mathbf{g}\rightarrow\mathbf{z}_1\rightarrow\mathbf{z}_2\)). Outputting the entire \(V\) from a single MLP is impractical for large number of modes, and the presented architectures instead partition \(V\) into subsets and then use a separate, smaller MLP for each subset, \(V_i = \textrm{MLP}_i(\mathbf{z}_2)\). Complex \(\mathbf{h}\) values are handled in \(\sigma\) by \(\mathbf{h}\rightarrow\textrm{cat}[|\mathbf{h}|, \textrm{Re}(\mathbf{h}), \textrm{Im}(\mathbf{h})]\). No learned or prescribed residual gating mechanism is used in the presented architectures for simplicity, but the iterative refinement procedure still applies: an initial output \(\hat V_1\) can be applied to \(U_1\) within \(A_i\) to "deform" \(U_1\) toward \(U_2\). Inference is then performed using the updated \(A_i\) matrices, and the estimate composes as \(V = \hat V_2\hat V_1\), \(\hat U_2 = V U_1\). Two inference passes are applied during training and testing.

In practice, illumination and detection involve distinct optical systems and each imposes its own transformation on the measurement, so what the architecture sees is \(BA_iC\) rather than \(A_i\), where the transformations \(B\) and \(C\) are fixed properties of the system rather than of the fiber \cite{bharadwaj2026}. These transformations break the reciprocity relation of Eq.~\eqref{eq:TM} but they do not obstruct recovery since each feature \(h^i_{jk}\) is a bilinear form in the measurements and the filters are mutable: for invertible \(B\), \(C\) and synthetically trained filters \(\bm\psi\), \(\bm\phi\), the experimentally trained (or fine-tuned) filters would become \((B^{-1})^{\dagger}\bm\phi\) and \(C^{-1}\bm\psi\), leaving \(h^i_{jk}\) intact (experimental filters for non-invertible transformations would instead move towards the available orthogonal projection, \(h_{jk}^i\) would change, and downstream weights would likewise need to be fine-tuned).

More generally speaking, the encoder does not commit to the basis in which the measurement is expressed: \(A_i\) is an operator and its rank is a basis-free property, thus the set of features the encoder can compute is unchanged under any invertible change of basis on either side of the measurement. This distinguishes it from the locality and translation structure that image-domain architectures impose on a pixel basis that has no counterpart in a measurement whose content is mode coupling.

\section{Results}

We trained our architecture under the protocol described in Table \ref{tab:training_protocol} on a synthesized DRAKA WideCap OM5 50 \(\mu\)m core fiber at \(\lambda =\) 532 nm with both circular-polarization states (\(\sigma_{+},\sigma_{-}\)) and \(N=300\) modes per polarization (600 total modes). We trained three separate models with varying levels of zero-mean gaussian noise, \(\sigma = \{0.0,0.04,0.16\}\), added to \(\{M_i\}\). We also trained a model on a synthesized step-index fiber as a reference with \(N=303\) using the Linearly Polarized basis and zero-mean \(\sigma=0.02\) gaussian noise added to its \(\{M_i\}\). The evaluation metric is fiber fidelity \(F = |\mathrm{tr}(V_{\mathrm{pred}}^{\dagger} V_\mathrm{gt})|/(\|V_{\mathrm{pred}}\|_F\|V_\mathrm{gt}\|_F)\) where \(V_\mathrm{pred}\) is the model output and \(V_\mathrm{gt}\) is the ground truth matrix.

\begin{figure}[h!]
  \centering
  \includegraphics[width=\columnwidth]{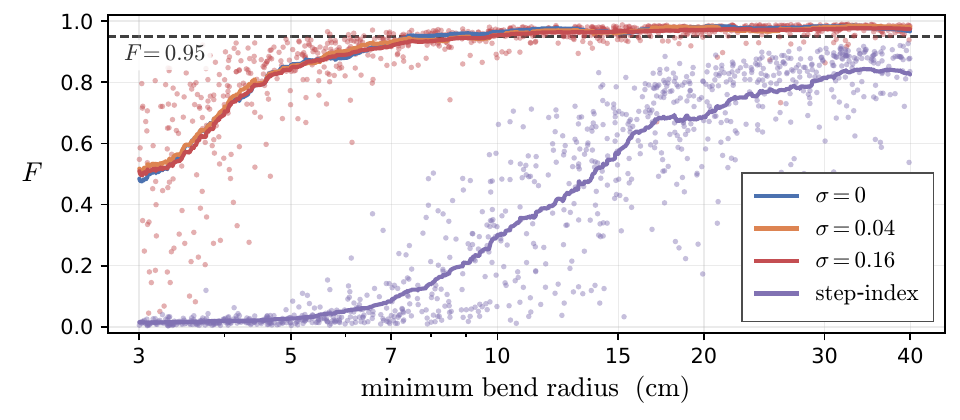}\\[4pt]
  \includegraphics[width=\columnwidth]{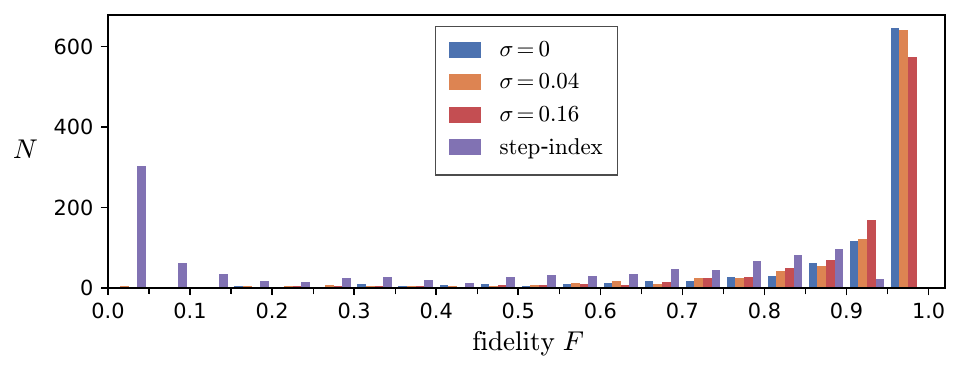}
  \caption{Recovery fidelity on the thousand held-out fibers for the three GRIN models trained at detector noise \(\sigma = 0\), \(0.04\) and \(0.16\), and the step-index model. each evaluated at its own \(\sigma\). Top: Scatter plot of \(F\) as a function of the minimum bend radius, with a rolling mean over \(80\) fibers. Individual scatter points are shown for the \(\sigma = 0.16\) model. Bottom: Histogram of the same fidelities.}
  \label{fig:sigma}
\end{figure}

\begin{table}[h!]
  \centering
  \caption{Fidelity recovery for the \(\sigma=0.16\) GRIN model on held-out testing data with mean initial fidelity \(F=|\mathrm{tr}(U_2 U_1^\dagger)|/(\sqrt{N}\,\|U_2 U_1^\dagger\|_F) = 0.1816\).}
  \vspace{0.5cm}
  \label{tab:fid_regime_vector}
  \begin{tabular}{l cccc}
    \toprule
    Subset & $n$ & Mean $F$ & $p_{5}$ & Worst \\
    \midrule
    All held-out fibers                    & 1000 & 0.897 & 0.551 & 0.045 \\
    Min bend radius \(\geq 5\,\)cm         &  808 & 0.953 & 0.869 & 0.603 \\
    Min bend radius \(\geq 7\,\)cm         &  692 & 0.964 & 0.919 & 0.671 \\
    Min bend radius \(\geq 10\,\)cm        &  566 & 0.970 & 0.934 & 0.671 \\
    \bottomrule
  \end{tabular}
\end{table}

\begin{table}[h]
  \centering
  \caption{Training protocol}
  \label{tab:training_protocol}
  \begin{threeparttable}
  \begin{tabular}{ll}
    \toprule
    Parameter & Value \\
    \midrule
    Training deformations & 19{,}000 \\
    Testing deformations  & 1000 \\
    Loss\tnote{1}         & $\min\limits_{\theta}\|Q(\theta)V_{\mathrm{pred}}-V\|_F^2/\|V\|_F^2$ \\
    Optimizer             & AdamW \\
    Learning rate         & $3\times10^{-4}$ \\
    Schedule              & cosine annealing to $0$ \\
    Epochs                & 40 \\
    Batch size            & 24 \\
    Inference passes      & 2 \\
    \bottomrule
  \end{tabular}
  \begin{tablenotes}[flushleft]\footnotesize
    \item[1] In the circular polarization basis
      \(Q(\theta)=\mathrm{Diag}(e^{i\theta}I_N,\,e^{-i\theta}I_N)\),
      where \(\theta\) is determined analytically.
  \end{tablenotes}
  \end{threeparttable}
\end{table}
\subsection{Learned Curvature Encoding}

\begin{figure}[h!]
  \centering
  \includegraphics[width=0.70\columnwidth]{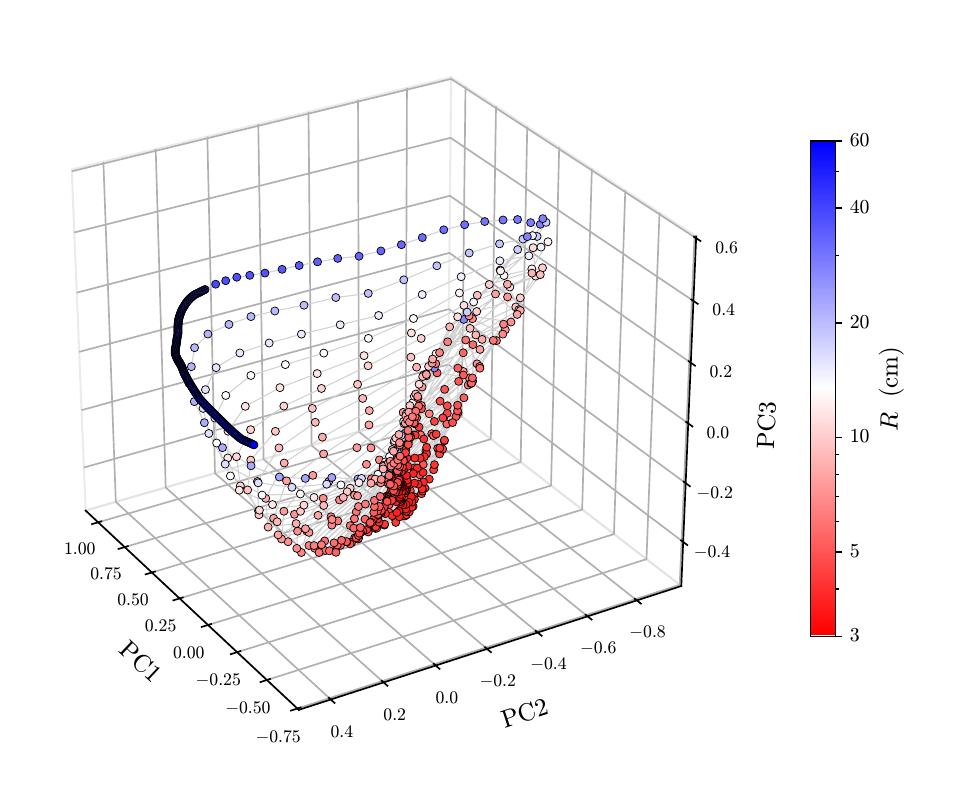}\\[4pt]
  \includegraphics[width=0.70\columnwidth]{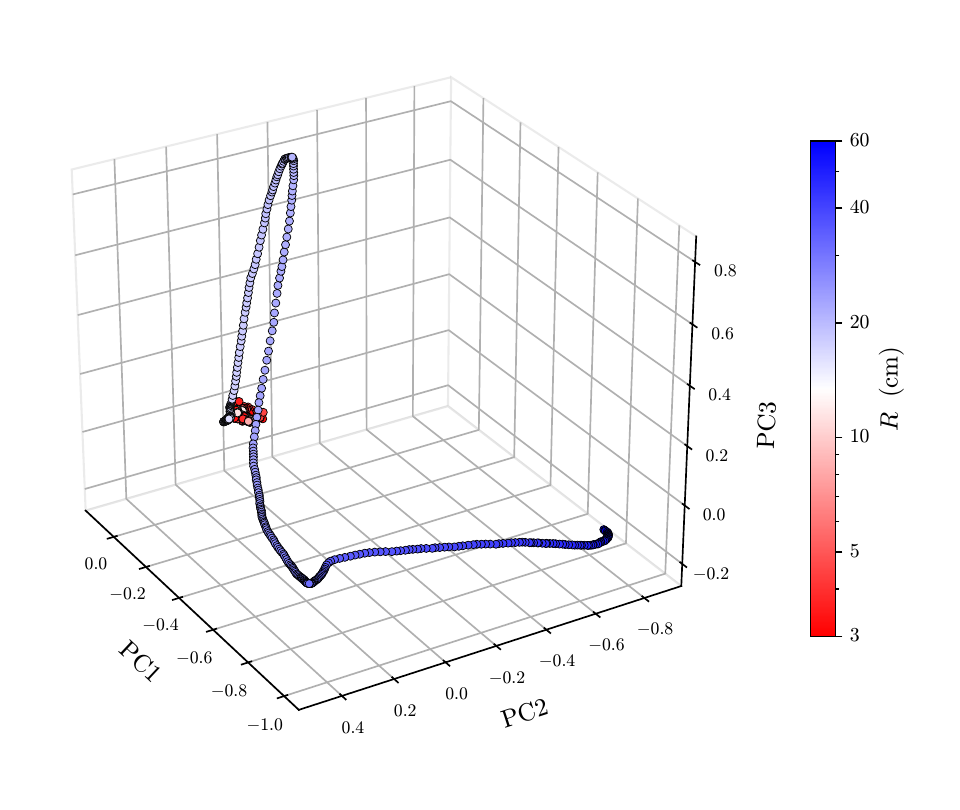}
  \caption{The three largest principal components of \(\mathbf{z}_2\) for the initial 620 wound fibers plus an additional 200 fibers in the \(R \in [40,60]\) cm range for the \(\sigma=0.16\) GRIN fiber (top) and step-index fiber (bottom), colored by radius of curvature. Gray lines indicate trajectory path. The architectures never saw bends with \(R\geq 40\) cm, so that curvature regime is outside of the training distribution but still traverses the manifold of both architectures as if it was in the distribution and demonstrates generalized capabilities. However, curvature information is much better compressed in the GRIN encoder, maintaining low-dimensional structure until about \(R=7\) cm, whereas low-dimensional representation in the step-index encoder breaks down near \(R=20\) cm.}
  \label{fig:manifold}
\end{figure}

To examine how the architecture organizes measurement information from these arbitrarily deformed GRIN fibers, we constructed 820 fibers each wound into a single circle of constant curvature, with radii of curvature between 3 and 60 cm. While these fibers are unphysical (many of them wind into themselves), they isolate a single scalar deformation parameter and allow it to be swept continuously. Furthermore, they appear in neither the training nor the testing set, and radii of curvature larger than 40 cm are never seen by the trained models. Figure \ref{fig:manifold} shows the main MLP's second hidden layer activations \(\mathbf{z}_2\) for each fiber projected onto the first three principal components. The architecture only sees the three measurements \(\{M_i\}\), and plotting these projections as a function of curvature shows the parameter traversing a three-dimensional manifold within the model's latent space. The first two components correspond to a local phase and the third component corresponds to a coarse global structure. The encoded curvature loses its structure around \(R=7\) cm, in agreement with the testing dataset results, and provides one possible explanation why some stray fibers with large \(R\) in Figure \ref{fig:sigma} were poorly recovered: their latent vector sits near the region with broken structure and was incorrectly decoded. The step-index model, despite the exact same encoder, could not encode the curvature as well, which is in agreement with \cite{flaes2018}. 

\begin{figure}[h!]
  \centering
  \includegraphics[width=1.0\columnwidth]{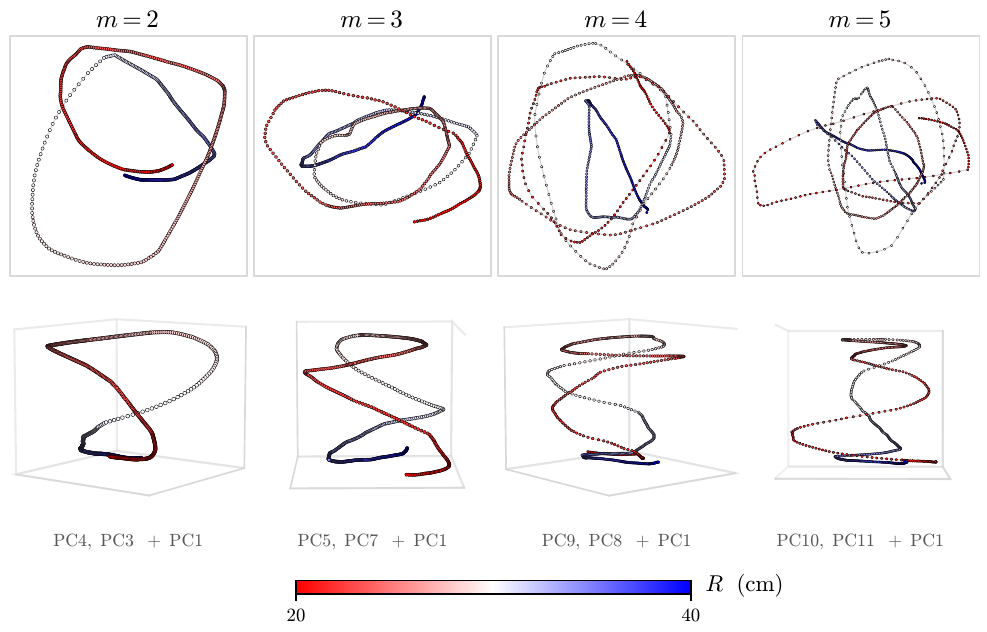}
  \caption{Curvature of the \(\sigma=0.16\) GRIN model encoded in \(\mathbf{z}_2\) as higher-ordered harmonics.}
  \label{fig:harmonics}
\end{figure}

We then looked to see if curvature is encoded by the model in higher order harmonics within \(\mathbf{z}_2\) with a dense sampling of the same constant curvature fibers from \(R=40\) cm to \(R=20\) cm. Figure \ref{fig:harmonics} shows that curvature is indeed very clearly encoded in higher order harmonics, and extents until the 12-14th harmonic for this range of \(R\).

\subsection{Learned Relative Positional Encoding}

\begin{figure}[h!]
    \centering
    \begin{subfigure}{\linewidth}
        \centering
        \includegraphics[width=0.85\linewidth, trim=78.9 38.6 47.5 23.6, clip]{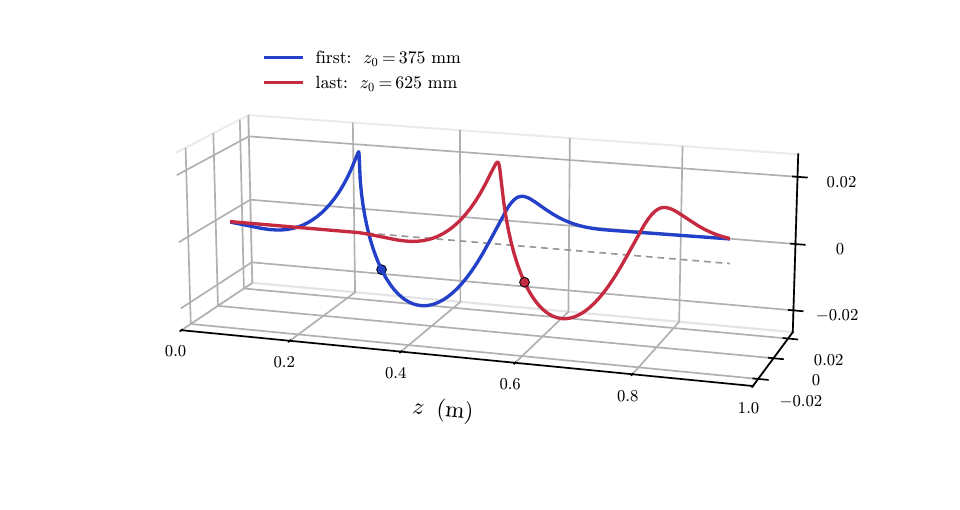}
        \caption{The first (\(z_0 = 375\) mm) and last (\(z_0 = 625\) mm) of the 10{,}800 fibers, where \(z_0\) (indicated by the dot) is the center of the spiral deformation.}
        \label{fig:gabor-ablation-geometry}
    \end{subfigure}

    \vspace{6pt}
    \begin{subfigure}{\linewidth}
        \centering
        \includegraphics[width=0.75\linewidth, trim=26.5 49.1 12.8 54.3, clip]{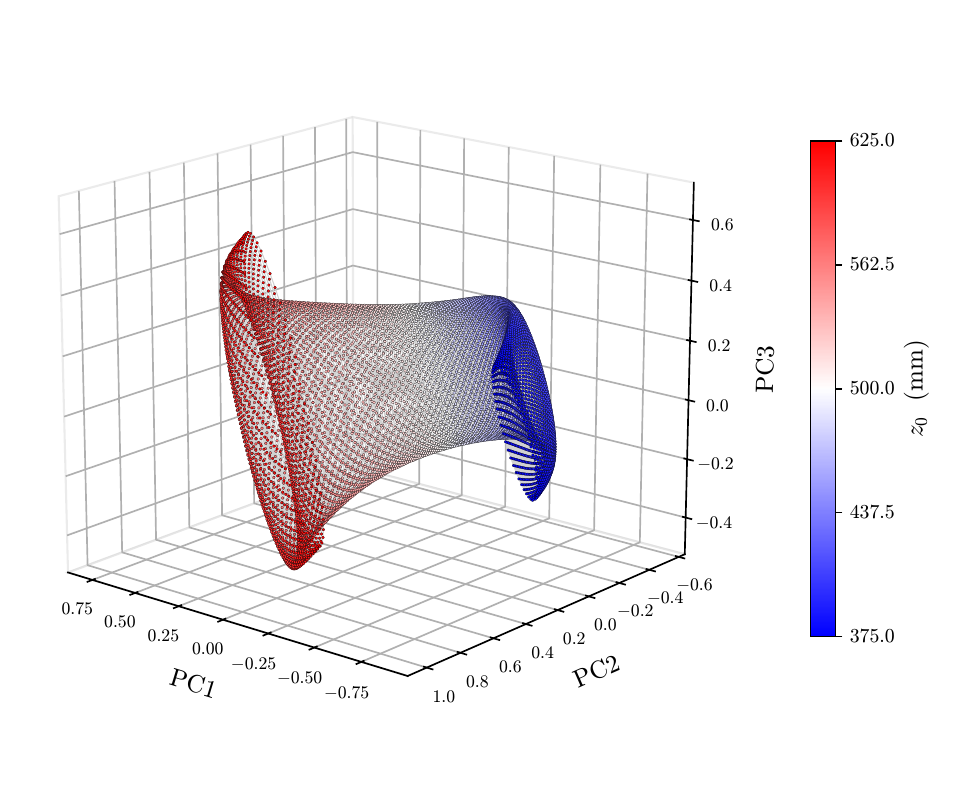}
        \caption{The three largest principal components of \(\textbf{z}_2\) for all 10{,}800 fibers, colored by \(z_0\).}
        \label{fig:gabor-ablation-pca}
    \end{subfigure}
    \caption{A localized spiral deformation translated along the middle quarter of the fiber. None of these fibers appear in the training or testing sets.}
    \label{fig:gabor-ablation}
\end{figure}

Another encoded property is the location of a particular deformation along the fiber. For a fiber with a localized defect that is partitioned into \(n_\text{total}\) sections, the TM can be represented by 
\begin{equation}
    U(n) = \left(e^{-iH_0\Delta z}\right)^{m-n}U_\text{defect}\left(e^{-iH_0\Delta z}\right)^{n},
\end{equation}
where there are \(m\) straight fiber segments, the defect is described by the remaining \(n_\text{total}-m\) segments, \(n\) indicates where the defect exists along the fiber, and \(H_0\) is the unbent Hamiltonian. Changing \(n\) corresponds to changing \(X\) in Eq. \eqref{eq:swap}, and to see if the model learned to encode this relational structure, we constructed 10{,}800 fibers in which the same localized spiral deformation (complex Gabor wavelet) is translated along the middle quarter of the fiber (Figure \ref{fig:gabor-ablation-geometry}). Moving the defect conjugates the TM by straight propagation, \(U(n) = e^{-iH_0 n\Delta z}\,U(0)\,e^{iH_0 n\Delta z}\), so each element only acquires a phase \(e^{-i(\beta_j-\beta_k)n\Delta z}\), where \(\beta_j\) are the propagation constants, and position enters the TM in the same way it enters a rotary positional encoding.

Figure \ref{fig:gabor-ablation-pca} shows that the architecture learned this structure. The latent \(\mathbf{z}_2\) oscillates with a period within 0.4\% of \(2\pi/\Omega\), and over a few millimeters its two largest principal components trace a circle, so the defect position is encoded as an angle while a slower component across the full sweep carries the coarse position along the fiber. Neither the defect position nor the self-imaging period is supplied to the architecture at any point, and like the curvature ablation, this relational structure is something it arrives at from the measurements alone.

\section{Discussion and Limitations}
The parameters of the simulated fiber and proximal-only system were chosen to go beyond proof-of-concept: the refractive index perturbations are the measured profiles of a Prysmian DRAKA WideCap OM5 50 \(\mu\)m core fiber and the Hamiltonian of Eq.~\eqref{eq:Hfull} contains the spin-orbit coupling together with all three birefringence contributions. The deformation regime spans minimum bend radii from 3 to 40 cm with out-of-plane displacement, non-adiabatic curvature changes, and twists. The straight calibration \(U_1\) retains a fidelity of 0.12-0.22 (mean 0.18) against the deformed \(U_2\), indicating that these are deformations under which a fixed calibration is unusable. The measurements are collected in a realistic, non-ideal manner where the fiber is perturbed independently during each probe signal (Eq. \eqref{eq:jitter}) so that no \(M_i\) corresponds to a single fiber state, and with system noise added on top of that (Figure \ref{fig:M3_noise}). Each of these choices moves the setting toward a working instrument and away from the regime in which recovery would be easiest.

Furthermore, our construction poses the hardest version of the recovery problem. Each deformation operator \(V\) is recovered in a single step between the straight calibration state \(U_1\) and a severely deformed \(U_2\), and nothing in the formulation requires this, since \(V = U_2U_1^\dagger\) holds for any previously known state in place of \(U_1\). A deployed endoscope deforms continuously, so the most recently recovered TM could and should serve as the reference where each update only needs to recover the incremental deformation accumulated over one time step, which is a far better conditioned problem than the full excursion from straight. From this perspective, we consider the fidelities in Table \ref{tab:fid_regime_vector} to be a lower bound on what the same architecture would achieve operating as a closed tracking loop with appropriate algorithms.

While the fibers are synthetic, the path from these results to an experimentally verified device does not require discarding the synthetic data. Every term in Eq.~\eqref{eq:Hfull} corresponds to an identified physical effect or fiber defect rather than a fitted degree of freedom: the parabolic potential and its measured perturbation, the bend-induced tilt, the twist rotation, the spin-orbit coupling, and the three birefringence contributions each have a separate physical origin, so a discrepancy between simulated and experimental measurements can be attributed to particular terms and corrected rather than absorbed into a global mismatch. The same holds for the measurement chain, where the sources we do not currently model are characterizable on a bench, and once characterized they enter the simulator in the same way that the acquisition jitter and system noise already do. Figure \ref{fig:sigma} shows that recovery is robust to measurement noise, with the \(\sigma=0.16\) measurements containing significant magnitude and phase errors, indicating that the architecture is not relying on artifacts particular to noise-free data. We therefore expect a practical route to an instrument to be pretraining on synthetic data made as faithful as the characterization allows, followed by fine-tuning on a comparatively small experimental set. 


\section{Conclusion}
In this work we introduce a bilinear neural network architecture for proximal-only TM recovery that achieves an average fidelity recovery of \(F=0.897\) on arbitrarily deformed GRIN MMFs with 600 guided modes (\(F\geq0.95\) when restricted to bending curvatures larger than 5 cm). We also perform several ablations demonstrating the learned model's ability to identify structure and generalize to fibers outside of the training distribution. The next step is to reproduce these capabilities on experimentally acquired measurements, and we believe the results here establish that the effort is warranted.

\bibliographystyle{unsrtnat}
\bibliography{refs}

\appendix

\section{Propagation Hamiltonian}
\label{app:hamiltonian}

\subsection{Scalar Approach}
The scalar field in a weakly guiding fiber obeys the Helmholtz equation \(\nabla^2 E + k_0^2 n^2(x,y,z)\,E = 0\). The ansatz \(E = \psi(x,y,z)\,e^{ikz}\) with \(k = n_0k_0\), the free-space wavenumber \(k_0 = 2\pi/\lambda\) and \(n_0\) the index on the fiber axis, gives

\begin{equation}
    \nabla^2\psi + 2ik\,\partial_z\psi + \left(k_0^2n^2 - k^2\right)\psi = 0 .
\end{equation}

When \(\psi\) varies slowly w.r.t \(z\) on the scale of a wavelength, the approximation \(|\partial_z^2\psi| \ll |2k\,\partial_z\psi|\) can be made and the \(\partial_z^2\psi\) term can be neglected. What remains is a 2-D Schr\"odinger-like equation on the \(xy\)-plane in which the propagation coordinate \(z\) plays the role of time,
\begin{equation}
    i\,\partial_z\psi = H\psi, \qquad
    H = \frac{p^2}{2k} + V(\bm r), \qquad
    V = \frac{k}{2}\left(1 - \frac{n^2}{n_0^2}\right),
    \label{eq:paraxial}
\end{equation}
with \(\bm p = -i\nabla_\perp\) acting on the two transverse (\(x\), \(y\)) coordinates only. For the ideal parabolic profile \(n^2(r) = n_0^2\big[1 - 2\Delta\,(r/a)^2\big]\), with core radius \(a\) and

\begin{equation}
    \Delta = \frac{n_0^2 - n_{\mathrm{cl}}^2}{2n_0^2} = \frac{\mathrm{NA}^2}{2n_0^2}
    \label{eq:contrast}
\end{equation}

is the relative index contrast between the axis and the cladding, Eq.~\eqref{eq:paraxial} gives a harmonic potential,
\begin{equation}
    V_0 = \frac{k}{2}\,\Omega^2 r^2, \qquad \Omega = \frac{\sqrt{2\Delta}}{a},
\end{equation}
and the unperturbed problem is the 2-D isotropic oscillator, whose eigenmodes are the Laguerre--Gauss modes \(\ket{\ell,m}\) with propagation constants \(\beta_g\) for each mode group index \(g = |\ell| + 2m\).

Bending enters through the geometry. A fiber bent to a local radius of curvature \(R\) is equivalent to a straight fiber whose index profile is tilted across the core \cite{heiblum1975, flaes2018}. To first order in \(r/R\) the equivalent profile is \(n_{\mathrm{eq}}^2 \approx n^2\left(1 + 2\,\bm\kappa\!\cdot\!\bm r\right)\) where \(\bm\kappa\) is the curvature vector with magnitude \(1/R\) pointing along the bend normal, and substituting into Eq.~\eqref{eq:paraxial} adds the term \(-k\,\bm\kappa\!\cdot\!\bm r\), linear in the transverse coordinate. The bend also compresses the glass on the inside and stretches it on the outside, changing the density and with it the index. Writing \(n-1\) as proportional to density and using \(\varepsilon_{xx} = \varepsilon_{yy} = -\sigma\,\varepsilon_{zz}\), where \(\sigma\) is the Poisson ratio, that correction leaves the form of the term alone and rescales the curvature itself \cite{ploschner2015},
\begin{equation}
    \eta = 1 - (1 - 2\sigma)\,\frac{n_0 - 1}{n_0} ,
    \label{eq:eta}
\end{equation}
so the fiber bends as though its curvature were \(\eta\bm\kappa\). For fused silica \(\sigma = 0.17\), giving \(\eta = 0.786\) at \(n_0 = 1.4783\), against the \(0.77 \pm 0.02\) measured in \cite{ploschner2015}. The bend term is therefore
\begin{equation}
    V_{\mathrm{bend}}(z) = -k\,\eta\,\bm\kappa(z)\!\cdot\!\bm r .
    \label{eq:Vbend}
\end{equation}

Twist is a cross-section that rotates along the fiber \cite{snyderlove}, through an accumulated angle \(\theta(z)\) at rate \(\tau(z) = d\theta/dz\), and collecting all contributions gives the ideal-fiber Hamiltonian
\begin{equation}
    H_{\mathrm{ideal}}(z) = \frac{p^2}{2k} + \frac{k\Omega^2}{2}\,r^2 - k\,\eta\,\bm\kappa(z)\!\cdot\!\bm r - \tau(z)\,L_z .
    \label{eq:Happ}
\end{equation}

Realistically, a manufactured fiber is not an ideal parabola. Its measured index deviates from the parabolic profile by the perturbation \(\delta n^2(x,y)\), and substituting \(n^2 = n_0^2\big[1 - 2\Delta\,(r/a)^2\big] + \delta n^2\) into Eq.~\eqref{eq:paraxial} leaves the harmonic potential untouched and adds the perturbative term,
\begin{equation}
    \delta V(\bm r) = -\frac{k}{2n_0^2}\,\delta n^2(x,y).
    \label{eq:defect}
\end{equation}
Crucially, this term breaks the degeneracy of the mode groups: the ideal parabola assigns a single \(\beta_g\) to every state of a given \(g\), and \(\delta V\) non-trivially perturbs each mode's propagation constant from the group value.

\subsection{Vector Approach}
GRIN fibers are not currently manufactured well enough for the scalar, single-polarization approach to be valid \cite{cao2023, turtaev2025}: ellipticity, refraction perturbations, and stresses introduced during the fiber draw couple \(\bm E\)'s polarization states and should be accounted for. The E-field obeys the vector wave equation \cite{snyderlove}
\begin{equation}
    \nabla^2\bm E + k_0^2n^2\bm E
    + \nabla\!\left(\bm E\!\cdot\!\nabla\ln n^2\right) = 0 .
    \label{eq:vector}
\end{equation}
Using the two-state Jones vector \(\psi\) \cite{jones1941} in the E-field ansatz, the scalar Hamiltonian in Eq. \eqref{eq:paraxial} carries over unchanged but the last term of Eq. \eqref{eq:vector} introduces the spin-orbit coupling term \(H_\mathrm{SO}\), 

\begin{equation}
    H_{\mathrm{SO}} = -\frac{1}{4k}
    \begin{pmatrix}
        \partial_{-}\,w_+ & \partial_{-}\,w_{-} \\[2pt]
        \partial_+\,w_+     & \partial_+\, w_{-}
    \end{pmatrix},\qquad \partial_{\pm} = \partial_x \pm i\partial_y , \quad
    w_{\pm} = \partial_{\pm} \ln n^2 ,
    \label{eq:so}
\end{equation}

where \(\psi\) is projected onto \(\bm e_\pm = (\bm e_x \pm i\bm e_y)/\sqrt2\) because the architecture assumes circular polarization.

Material anisotropy contributes three further terms acting on the polarization index. (1) Frozen stresses leave a fixed linear birefringence \cite{kaminow1981} whose slow axis sits at an angle \(\alpha\) set by the quadrupole component of the measured profile:

\begin{equation}
H_1 = \frac{b_1}{2}\big(e^{-2i\alpha}\sigma_+ + e^{2i\alpha}\sigma_-\big),
\end{equation}

where \(\sigma_\pm = (\sigma_x \pm i\sigma_y)/2\). (2) Bending compresses the glass on the inside of the bend and stretches it on the outside. The resulting stress-optics birefringence \cite{ulrich1980} has its principal axes set by the bend plane and its magnitude is quadratic in curvature. Define \(\phi(z)\) to be the angle \(\bm\kappa(z)\) makes with the \(x\)-axis of the co-rotating spatial profile,

\begin{equation}
H_2 = \frac{b_{2}\kappa^2}{2}\left(e^{-2i\phi}\sigma_+ + e^{2i\phi}\sigma_-\right),
\end{equation}

where \(\kappa = |\bm\kappa|\). This is identical to \(H_1\) with the bend plane in place of the frozen axis and \(b_2\kappa^2\) in place of \(b_1\) with the sign of \(b_2\) fixing which of the two axes is slow.
(3) Twisting shears the glass, and the photoelastic response to that shear is a circular birefringence 
\begin{equation}
    H_3 = b_3\,\tau(z)\,\sigma_z.
\end{equation}

Collecting every contribution gives the Hamiltonian used for our synthesized fibers,
\begin{equation}
H(z) = \frac{p^2}{2k} + \frac{k\Omega^2}{2}r^2 + \delta V(\bm r)
              - k\,\eta\,\bm\kappa(z)\!\cdot\!\bm r - \tau(z)L_z
              + H_{\mathrm{SO}} + H_1 + H_2 + H_3.
\label{eq:Hfull}
\end{equation}

\section{Fiber Model and Dataset Synthesis}
\label{app:simulator}

\subsection{Index profile and mode basis}
We simulate the Prysmian DRAKA WideCap OM5 fiber, with a 50\,\(\mu\)m core and NA 0.200, at \(\lambda = 532\) nm. The index profile is the ideal parabolic profile with the measured perturbation \(\Delta n^2(x,y)\) reported in  \cite{turtaev2025}.

\begin{table}[h!]
\centering
\caption{Simulated DRAKA fiber parameters.}
\label{tab:params}
\small
\begin{tabular}{llll}
\toprule
Symbol & Meaning & Value & Source \\
\midrule
\multicolumn{4}{l}{\textit{Fiber (Prysmian DRAKA WideCap OM5)}}\\[2pt]
\(\lambda\)   & build wavelength      & \(532\,\mathrm{nm}\)              & N.A.  \\
\(n_0\)       & index on axis         & \(1.4783\)                        & manufacturer \\
\(\mathrm{NA}\) & numerical aperture & \(0.200\)                         & manufacturer \\
\(a\)         & core radius           & \(25\,\mu\mathrm{m}\)           & manufacturer \\
\(R_{\mathrm{cl}}\) & cladding radius & \(62.5\,\mu\mathrm{m}\)         & manufacturer \\
\(\Delta\)    & index contrast        & \(9.15\times10^{-3}\)             & Eq.~\eqref{eq:contrast} \\
\(L\)         & fiber length          & \(1.0\,\mathrm{m}\)                & N.A. \\
\(k_0\)       & vacuum wavenumber & \(1.181\times10^{7}\,\mathrm{m^{-1}}\) & \(2\pi/\lambda\) \\
\(k\)         & axial wavenumber      & \(1.746\times10^{7}\,\mathrm{m^{-1}}\) & \(n_0k_0\) \\
\(\Omega\)    & oscillator frequency  & \(5412\,\mathrm{m^{-1}}\)         & \(\sqrt{2\Delta}/a\) \\
\(\sigma\)    & Poisson ratio         & \(0.17\)                         & fused silica \cite{ploschner2015} \\
\(\eta\)      & elasto-optic curvature factor & \(0.786\)                & Eq.~\eqref{eq:eta} \\
\addlinespace
\multicolumn{4}{l}{\textit{Polarization coefficients}}\\[2pt]
\(\alpha\)    & frozen-stress axis    & \(+18.5^\circ\)                   & \cite{turtaev2025}, quadrupole of \(\delta n^2\) \\
\(b_1\)       & frozen linear biref.  & \(3.0\,\mathrm{rad\,m^{-1}}\)     & no source \\
\(b_2\)       & bend stress-optic     & \(6.25\times10^{-3}\,\mathrm{rad\,m}\) & \cite{ulrich1980} \\
\(b_3\)       & twist activity & \(0.14\)                         & \cite{ulrich1979}; \(0.13\)--\(0.16\) \\
\bottomrule
\end{tabular}
\end{table}

\begin{table}[h!]
\centering
\caption{Step-index fiber parameters. The fiber is mode-matched to the DRAKA fiber of Table~\ref{tab:params}: same core radius, same \(n_0\), same wavelength, same defect field \(\delta n^2\), same polarization coefficients, and similar number of guided modes.}
\label{tab:params_step}
\small
\begin{tabular}{llll}
\toprule
Symbol & Meaning & Value & Source \\
\midrule
\multicolumn{4}{l}{\textit{Fiber (step-index, mode-matched to the DRAKA fiber)}}\\[2pt]
\(\lambda\)   & build wavelength      & \(532\,\mathrm{nm}\)              & matched \\
\(n_0\)       & core index            & \(1.4783\)                        & matched \\
\(n_{\mathrm{cl}}\) & cladding index  & \(1.47131\)                       & Eq.~\eqref{eq:contrast} \\
\(\mathrm{NA}\) & numerical aperture  & \(0.1436\)                        & chosen for \(452\) guided modes \\
\(a\)         & core radius           & \(25\,\mu\mathrm{m}\)             & matched \\
\(R_{\mathrm{cl}}\) & cladding radius & \(62.5\,\mu\mathrm{m}\)           & matched \\
\(\Delta\)    & index contrast        & \(4.72\times10^{-3}\)             & Eq.~\eqref{eq:contrast} \\
\(V\)         & normalized frequency  & \(42.40\)                         & \(k_0 a\,\mathrm{NA}\) \\
\(L\)         & fiber length          & \(1.0\,\mathrm{m}\)               & matched \\
\(k_0\)       & vacuum wavenumber     & \(1.181\times10^{7}\,\mathrm{m^{-1}}\) & \(2\pi/\lambda\) \\
\(k\)         & axial wavenumber      & \(1.746\times10^{7}\,\mathrm{m^{-1}}\) & \(n_0k_0\) \\
\(\Omega_{\mathrm{eff}}\) & oscillator frequency & \(3048\,\mathrm{m^{-1}}\)   & \(\sqrt{2\Delta}/a\) \\
\(\sigma\)    & Poisson ratio         & \(0.17\)                          & fused silica \cite{ploschner2015} \\
\(\eta\)      & elasto-optic curvature factor & \(0.786\)                 & Eq.~\eqref{eq:eta} \\
\addlinespace
\multicolumn{4}{l}{\textit{Polarization coefficients (unchanged from Table~\ref{tab:params})}}\\[2pt]
\(\alpha\)    & frozen-stress axis    & \(+18.5^\circ\)                   & matched \\
\(b_1\)       & frozen linear biref.  & \(3.0\,\mathrm{rad\,m^{-1}}\)     & matched \\
\(b_2\)       & bend stress-optic     & \(6.25\times10^{-3}\,\mathrm{rad\,m}\) & matched \\
\(b_3\)       & twist activity        & \(0.14\)                          & matched \\
\bottomrule
\end{tabular}
\end{table}

\subsection{Deformation synthesis}
Each deformation is generated by partitioning the fiber into \(n\) equal-length sections, where \(n\) is uniformly sampled from \{1,2,3,4,5,6\}. Each section is assigned a curvature vector \(\bm\kappa\) whose magnitude corresponds to a radius of curvature drawn uniformly from 3 to 40 cm and whose direction is drawn uniformly on the circle, so that the deformation is not confined to a single plane. The resulting piecewise-constant \(\bm\kappa(z)\) is convolved with a Gaussian kernel to produce a continuous curvature profile, which makes the curvature vary smoothly along the fiber while remaining non-adiabatic. A twist \(\tau(z)\) is applied over the full length, with a total accumulation uniformly sampled from \([-2\,\textrm{rad}, 2\,\textrm{rad}]\).

\begin{figure}[h!]
    \centering
    \includegraphics[width=\textwidth]{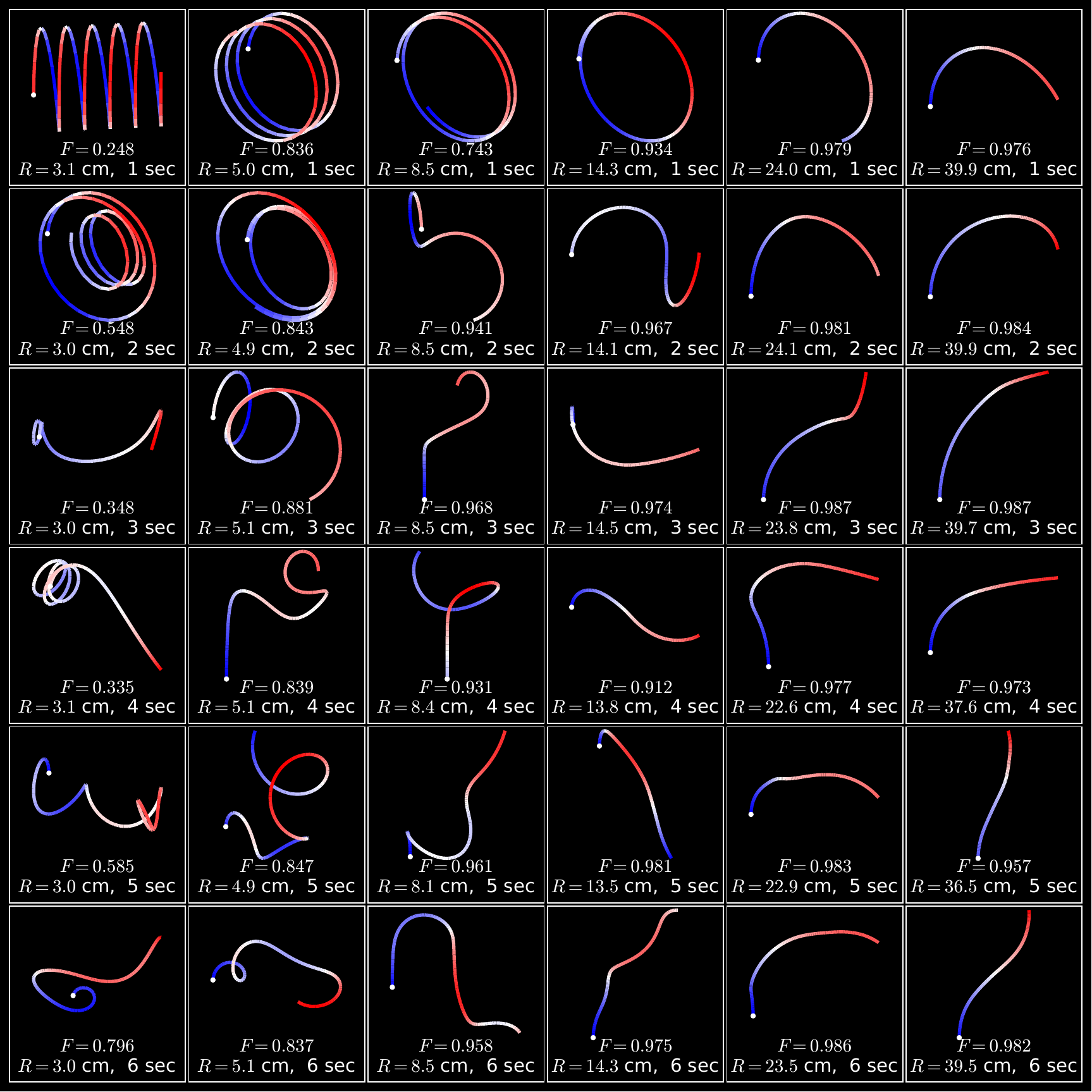}
    \caption{Generated fiber deformations in the held-out test set.}
    \label{fig:test_fibers}
\end{figure}

\subsection{Transmission matrix construction}
The smoothed deformation profile is discretized into 100 segments of constant \(\bm\kappa\) and \(\tau\), the Hamiltonian in Eq.~\eqref{eq:Hfull} is evaluated on each segment in the retained mode basis, and the transmission matrix is the ordered product of segment propagators as in Eq.~\eqref{eq:segments}. The undeformed transmission matrix \(U_1\) is obtained from the same construction with \(\bm\kappa = 0\) and \(\tau = 0\), so that \(U_1\) carries the fiber's defects.

\subsection{Reflector stack and measurement acquisition}
The distal stack consists of a straight partial reflector \(R_1 = \rho_1 I\), a tilted partial reflector \(R_2\), and a second tilted partial reflector \(R_3\) whose tilt axis makes an angle of \(\pi/4\) with that of \(R_2\). Both \(R_2\) and \(R_3\) are tilted \(0.6\deg\). Reflector \(R_2\) is simulated 3mm behind \(R_1\), and \(R_3\) is simulated 5mm behind \(R_2\). The transmission matrices \(T_i\) are the scalars \(t_i I\) that satisfy energy conservation.


Each measurement \(M_i\) is assembled from \(N\) linearly independent probe signals, and each probe is subject to an independent rigid misalignment of the launch. We model it to first order in the four generators \(J_a \in \{X,\, Y,\, P_x,\, P_y\}\) of transverse displacement and tilt, so that the measured matrix becomes
\begin{equation}
    M' = M + i\sum_a \varepsilon_a\, J_a M D_a + i\,M \sum_a J_a D_a , \qquad D_a = \operatorname{diag}(d_{j,a}) ,
    \label{eq:jitter}
\end{equation}
where \(d_{j,a}\) is a randomly sampled from a Gaussian independently for every probe \(j\), whose standard deviation \(\sigma_j\) is itself assigned per fiber, stratified over \(\sigma_j \in [0,\,0.10]\) so that the population spans the range with mean \(0.05\), and \(\varepsilon_a = (1,1,-1,-1)\) is the parity of each generator under the reciprocity flip \(P\). The \(\sigma_j\) varies for each synthetic fiber's \(\{M_i\}\) measurements, but \(U_1\) is constructed with a smaller \(\sigma_j = 0.01\) since care can be taken before training to ensure it's more accurate with techniques like multi-look averaging. The measurements are assumed single-look and therefore contain more "jitter" noise. Two terms appear because the same physical misalignment is seen on the launch and again on the return path, and the right-acting operator \(\sum_a J_a D_a\) is dense rather than diagonal. Zero-mean Gaussian noise of standard deviation \(\sigma\) is then added to the reflection measurements.

\begin{figure}[h!]
\centering
\includegraphics[width=\textwidth]{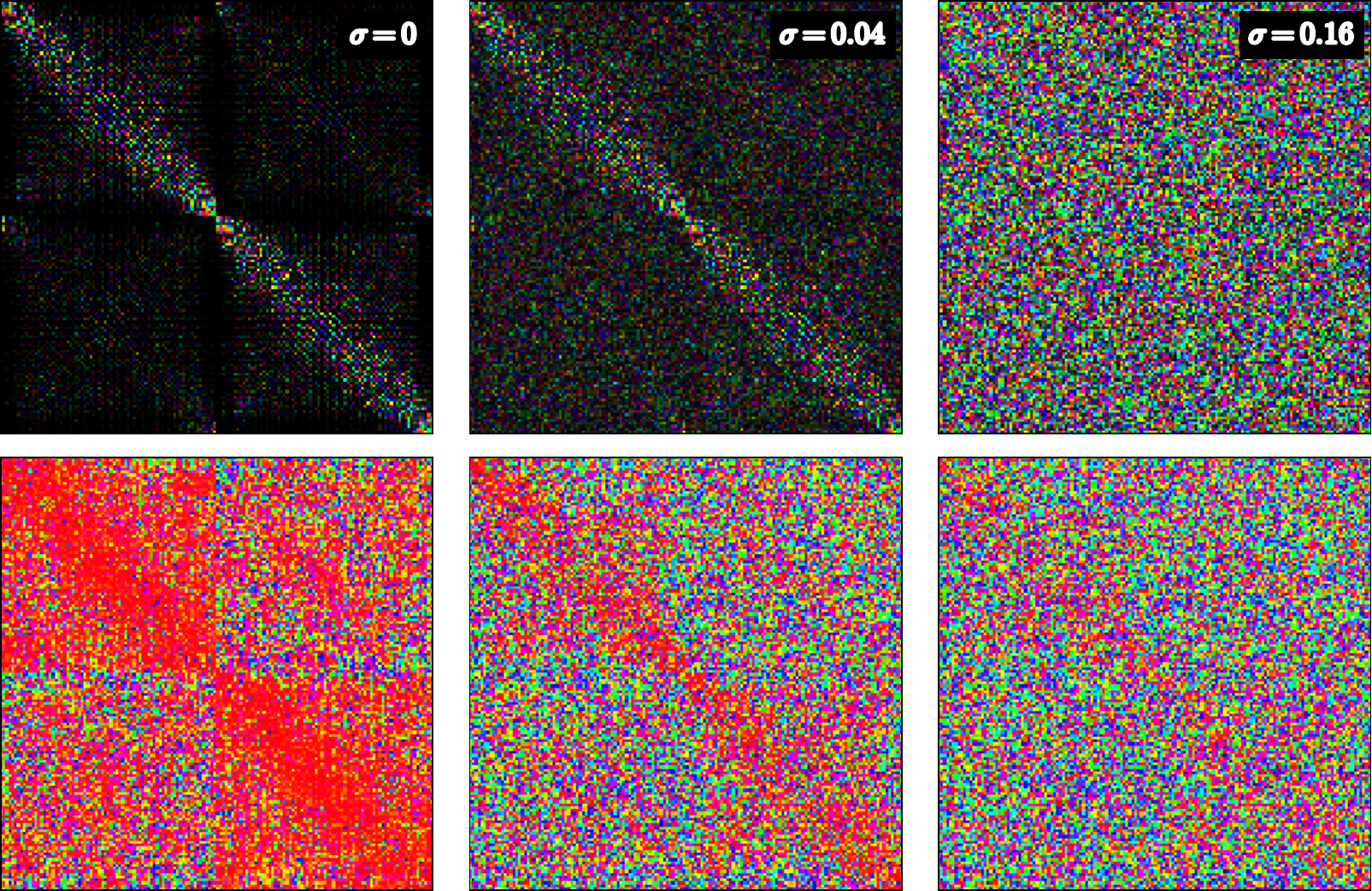}
\caption{The third reflector measurement \(M_3\) of a single deformed fiber at three noise levels. The top row is the measurement as the architecture receives it, where hue is phase and brightness is magnitude. The bottom row is the phase deviation of each matrix element from its ideal value, \(\arg(M^*_{\mathrm{ideal}}\circ M_\mathrm{measured})\), where \(\circ\) is the elementwise product and \(M^*_{\mathrm{ideal}}\) is the measurement that a single, unperturbed fiber state would produce without added noise. The \(\sigma=0\) column therefore isolates the per-probe deformation applied during acquisition, which is present in every column.}
\label{fig:M3_noise}
\end{figure}


\end{document}